\documentclass[USenglish,twocolumn]{article}

\ifx\directlua\undefined\ifx\XeTeXcharclass\undefined
  \usepackage[utf8]{inputenc}                           
  \else\RequirePackage[no-math]{fontspec}[2017/03/31]\fi 
  \else\RequirePackage[no-math]{fontspec}[2017/03/31]\fi 
\usepackage[sort&compress,square,numbers]{natbib}
\usepackage[big,online]{dgruyter}
\usepackage{pdfpages}

\theoremstyle{dgthm}

\newcommand{\abs}[1]{\left| #1 \right|}

\newcommand{\bra}[1]{\left\langle #1 \right|}
\newcommand{\ket}[1]{\left| #1 \right\rangle}
\newcommand{\braket}[2]{\left\langle {#1{\left| \vphantom{#1 #2} \right.} #2} \right\rangle}
\newcommand{\qo}[1]{``#1''}

\renewcommand{\epsilon}{\varepsilon}

\def\VR{\kern-\arraycolsep\strut\vrule &\kern-\arraycolsep}
\def\vr{\kern-\arraycolsep & \kern-\arraycolsep}
\usepackage{sistyle}

\theoremstyle{dgdef}

\begin{document}

\articletype{Research Article}
\author[1]{Tareq Jaouni}
\author[2]{Xuemei Gu} 
\author[2]{Mario Krenn} 
\author[1]{Alessio D’Errico}
\author[1,2,3]{Ebrahim Karimi}
\affil[1]{Nexus for Quantum Technologies, University of Ottawa, Ottawa, K1N 6N5, ON, Canada}
\affil[2]{Max Planck Institute for the Science of Light, Erlangen, Germany}
\affil[3]{Institute for Quantum Studies, Chapman University, Orange, California 92866, USA}

\title{Tutorial: Hong-Ou-Mandel interference with Structured Photons}
\runningtitle{Structured Photons's HOM}
\abstract{The Hong-Ou-Mandel (HOM) effect, an effective two-photon interference phenomenon, is a cornerstone of quantum optics and a key tool for linear optical quantum information processing. While the HOM effect has been extensively studied both theoretically and experimentally for various photonic quantum states, particularly in the spectral domain, detailed overviews of its behaviour for structured photons -- those with complex spatial profiles -- under arbitrary spatial mode measurement schemes are still lacking. This tutorial aims to fill this gap by providing a comprehensive theoretical analysis of the HOM effect for structured photons, including an arbitrary mode projection on quantum interference outcomes. The tutorial also provides analytical, closed-form expressions of the HOM visibility under different measurement conditions, which is a crucial contribution for its application in computational and artificial-intelligence-driven discovery of new quantum experiments exploiting the power of photons with complex spatial modes.}
\keywords{Two-photon interferometry; Hong-Ou-Mandel; Structured Photons; Orbital Angular Momentum; Laugerre Gauss Modes}
\journalname{Nanophotonics}

\journalyear{2024}
\journalvolume{aop}

\maketitle

\section{Introduction} 
With the advent of second quantisation, the study of optical elements such as mirrors, wave plates, and beamsplitters, along with their interactions with various quantum states, emerged as a natural domain for exploration. This theoretical framework enabled quantum physicists to rigorously describe and analyze how quantum states evolve under the influence of optical systems, providing deeper insights into quantum optics and discovering novel optical phenomena. An important example is the action of a lossless beamsplitter on photonic quantum states, which was extensively studied by several research groups in the late 1980s. Their theoretical analyses revealed that indistinguishable photons tend to \qo{bunch} together at the output ports of a lossless beamsplitter, a phenomenon now known as the Hong-Ou-Mandel (HOM) effect \cite{ou2007multi}. The experimental verification of this effect was later carried out by Hong, Ou, and Mandel \cite{hong1987measurement} and independently by Rarity and Tapster \cite{rarity1989fourth}. Utilising photons generated through spontaneous parametric down-conversion (SPDC), these experiments not only confirmed the photon bunching effect, but also demonstrated its practical utility in quantum metrology, achieving sub-picosecond temporal precision \cite{lyons2023fluorescence}. Recent advancements have improved these techniques, achieving sub-femtosecond time-resolution capabilities \cite{lyons2018attosecond}.

Single-particle interference raises fundamental questions regarding path determination and the visibility of interference, challenging classical notions of locality in the context of quantum systems, i.e., wave-particle duality. In contrast, multi-particle interference, manifested by the HOM effect, introduces additional layers of foundational complexity, revealing unique many-body quantum features. For instance, the unitary action of a 50:50 beamsplitter on two independent, indistinguishable photons ensures that the photons always emerge from the same output port. This phenomenon arises from destructive quantum interference between \emph{two particles} that eliminates the possibility of photons exiting through separate output ports, a universal property of lossless beamsplitters irrespective of their physical construction, whether dielectric or metallic.
The quantum state of the photons is critical in determining the interference outcomes. For instance, by engineering the quantum state to mimic fermionic behaviour, the photons exhibit anti-bunching, always exiting through different output ports--a unique contrast to their bosonic counterpart \cite{zhang2016engineering,gao2022manipulating}. Such manipulations not only highlight the versatility of quantum states but also emphasise the rich interplay between quantum statistics and optical systems.
Although a beamsplitter is inherently a linear optical element, it is a fundamental tool for facilitating interactions between multiple photons--a crucial device for quantum information processing. By changing the quantum state of one of the input photons, a beamsplitter (or polarising beamsplitter) can effectively control the output quantum states and the distribution of photons at its ports --- varying the reflection and transmission coefficients allows to control the distribution further. This functionality underpins the implementation of a controlled-NOT (CNOT) gate \cite{pittman2003experimental, gasparoni2004realization}, a cornerstone for constructing all-optical quantum circuits.

A single photon carries various degrees of freedom, including polarisation, frequency, and spatial modes. While polarisation spans a two-dimensional vector space, frequency and spatial modes are inherently unbounded, offering infinite-dimensional Hilbert spaces. As a result, the quantum states of two photons may vary across any of these degrees of freedom. When two photons encounter a beamsplitter, the resulting interaction depends critically on the overlap between their quantum wavefunctions \cite{bouchard2020two}. This overlap determines the two-photon interference at the beamsplitter’s output ports. However, the detection systems—such as detectors positioned at the exit ports—and measurement processes introduce further complexities, which require attention. Measurement schemes, driven by the detectors' sensitivity, can significantly reduce the adequate Hilbert space of the photons to the subspaces detected. For instance, selecting photons within a specific frequency domain by means of bandpass filters reduces the frequency Hilbert space, which can either enhance or diminish the indistinguishability of the photons' quantum states \cite{ou2007multi}. Similarly, bucket detectors, which are insensitive to spatial modes, effectively project onto mixed spatial modes, ignoring spatial-mode information altogether.
Numerous studies have theoretically and experimentally explored the effects of a beamsplitter on photons characterised by quantum states across different degrees of freedom, such as frequency \cite{zhang2021high}, polarisation \cite{kim2003experimental}, orbital angular momentum \cite{nagali2009optimal}, radial quantum numbers \cite{karimi2014exploring}, and vector modes \cite{gao2024vector}. Despite this progress, there remains a gap in the literature: a comprehensive tutorial addressing the interplay between beamsplitters, detection systems, and photons with differing spatial modes. Such a tutorial would be a critical resource for understanding and designing experiments where spatial-mode mismatches and their impact on quantum interference are central considerations.

This tutorial is organized as follows. In Section 2, we review the theory of multiphoton states, which can be in several different spatial modes. In Section 3, we develop the theory of HOM interference between two frequency-degenerate photons taking into account both their spatial mode content and measurement devices. We specialize the results into different examples that highlight how different measurement configurations allow to recover different information on the bunching photons. In Section 4, we derive an analytical formula for the scalar product of Laguerre-Gauss modes with different mode parameters, which can provide a useful speedup in calculating the outcome of free-space optical quantum networks that we outline in the application Section 5.

\section{Multimode quantum optics}

The second quantisation of electric and magnetic fields provides a quantum formalism for describing electromagnetic waves at the single quantum level, i.e., photons. In fact, a photon is defined as an excitation of the vacuum state $\ket{0}$, expressed as:
\begin{equation}\label{eq:EMfield}
    \hat{a}^\dagger\ket{0}=\ket{1},
\end{equation}
where $\hat{a}^\dagger$ and $\hat{a}$ denote the creation and annihilation operators for photons, respectively \cite{mandel1995optical, scully1997quantum}. These operators satisfy the relationships  $\hat{a}^\dagger\ket{n}=\sqrt{n+1}\ket{n+1}$ and $\hat{a}\ket{n}=\sqrt{n}\ket{n-1}$, with $\ket{n}$ representing the quantum state of $n$ photons. In the above notation, the photon's quantum state is not determined. A single photon possesses several degrees of freedom, such as polarisation, frequency (or time), and spatial modes (transverse position or wave vector). Accordingly, the mode annihilation and creation operators can be labeled with these quantum numbers, i.e., $\hat{a}_{\sigma,\omega,\textbf{q}}$, $\hat{a}^\dagger_{\sigma,\omega,\textbf{q}}$. Here, the polarization state $\sigma$ can take values of $\{+1,-1\}$, corresponding to the left- or right-handed polarisation states, $\omega\in\Re$ stands for the Hilbert frequency space, and $\textbf{q}\in\Re^2$ refers to the transverse components of the wavevector $\mathbf{k}=(\mathbf{q},\sqrt{\omega^2/c^2-\abs{\mathbf{q}}^2})$. One can also define creation and annihilation operators in the \qo{transverse position space}, $\mathbf{r}_\bot $ using the Fourier transformation: $\hat{a}_{\sigma, \omega,\mathbf{r}_\bot}:=\int d^2q \, \exp(-i \mathbf{q}\cdot\mathbf{r}_\bot )\hat{a}_{\sigma, \omega,\mathbf{q}}$.  Of course, a single photon described by a given wavefunction can exist in a coherent or incoherent superposition of these degrees of freedom \cite{mandel1995optical,garrison2008quantum}, such as:
\begin{eqnarray}\label{eq:structuredphotons}
    \ket{1}_{f}&=&\hat{a}_f^{\dagger}\ket{0}\cr
    &=&\sum_\sigma\int d\omega\int d^2r_\bot\, f_\sigma(\textbf{r}_\bot,\omega)\, \hat{a}^\dagger_{\sigma,\omega,\textbf{r}_\bot}\,\ket{0},\quad
\end{eqnarray}
where $f_\sigma(\textbf{r}_\bot,\omega)$ is the expansion coefficient function, which can be interpreted as the probability amplitude of detecting a photon with frequency $\omega$ and polarization $\sigma$ in the transverse position $\textbf{r}_\bot$. For example, the state of a monochromatic, left-handed single photon with a spatial profile given by the Laguerre-Gauss modes \cite{siegman1986lasers} can be expressed as:
\begin{eqnarray}\label{eq:onephotonLG}
    \ket{1}_{+1,p,\ell}&:=&\hat{a}_{+1,p,\ell}^{\dagger}\ket{0}\cr
    &=&\int d^2{r}_\bot\, \text{LG}_{p,\ell}(\textbf{r}_\bot;z)\,\hat{a}_{+1}^\dagger(\textbf{r}_\bot)\ket{0}.
\end{eqnarray}
Here $\hat{a}_{+1}^\dagger(\textbf{r}_\bot)$ is the creation operator that generates a left-handed circularly polarized photon at transverse position $\textbf{r}_\bot$ in the propagation plane specified by the distance $z$. The $\text{LG}_{p,\ell}(\textbf{r}_\bot;z)$ are the LG modes, given by the following expression in cylindrical coordinates $(\textbf{r}_\bot;z)=(\rho,\phi;z)$,
\begin{align}\label{eq:LG}
    \text{LG}_{p,\ell}(\rho,\phi,z)&=\braket{\rho,\phi,z}{p,\ell}\cr
    &=\sqrt{\frac{2\,p!}{\pi(p+|\ell|)!}}\,\left(\frac{1}{w(z)}\right)\,\left(\frac{\sqrt{2}\,\rho}{w(z)}\right)^{|\ell|}\cr
    &\times\exp{\left(-\frac{\rho^2}{w(z)^2}\right)}\text{L}_p^{|\ell|}\left(\frac{2\rho^2}{w(z)^2}\right)\cr
    &\times\!e^{i\left(\ell\phi+(2p+|\ell|+1)\arctan{(\frac{z}{z_0})}+\frac{k \rho^2}{2R(z)}\right)}.\qquad
\end{align}
Here, $w(z)=w_0(1+(z/z_0)^2)^{1/2}$, $R(z)=z(1+(z_0/z)^2)$, $z_0=1/2 k w_0^2$, $k$ and $w_0$ are the radius of the beam, the curvature of the beam, the Rayleigh range, the wave vector, and the radius of the beam at the waist, respectively, and $L_p^{|\ell|}(.)$ are the generalized Laguerre polynomials. Note that we have neglected the propagation phase factor of $\exp{(i(k z-\omega t))}$. The index $\ell$ is an integer and is associated with the helical wavefront structure $\exp(i\ell \phi)$ which implies that these are eigenmodes of the Orbital Angular Momentum (OAM) operator with eigenvalue $\ell\hbar$ \cite{allen1992orbital}. The index $p$ is instead a positive integer that counts the number of amplitude zeros in the radial direction \cite{karimi2014radial,plick2015physical}.

In order to determine Laguerre-Gaussian modes, in addition to radial and azimuthal indices $p,\ell$, the beam waist needs to be fixed. Assuming a fixed beam waist $w_0$, LG modes form a complete set of orthogonal bases that satisfy the relationships $(p,\ell|p',\ell')=\delta_{p,p'}\delta_{\ell,\ell'}$ and $\sum_{p,\ell}|p,\ell)(p',\ell'|=\hat{1}$ -- here we use `rounded' kets, $|\ldots )$ to distinguish vectors in the Hilbert space of square integrable functions $\mathcal{L}^2(\Re^2)$ from elements of the Fock space. It is important to note that two LG modes with different beam waist parameter or with different wavefront curvature, even if possessing different radial indices, are \emph{not} orthogonal.

In a similar fashion, the quantum states of two or more photons can be described in terms of their respective quantum properties. Depending on the generation, manipulation, configuration, or experimental setup, these photons may share identical or possess different quantum states. For instance, $N$ photons can be prepared in the same spatial, polarisation, or frequency modes (indistinguishable states) or in different modes (distinguishable states). This formalism can be employed to represent separable states, entangled states, and other multi-photon quantum phenomena. To highlight this point, we report here the expression of two cases of $N$ photon states. \newline

\noindent\emph{$N$ identical photons having the same polarisation state, and spatial modes:}
Using $\hat{a}^{\dagger}\ket{n}=\sqrt{n+1}\ket{n+1}$ and Eq. \eqref{eq:onephotonLG} we obtain
\begin{align}\label{eq:NiP}
    \ket{N}_{\sigma,p,\ell}&:=\frac{1}{\sqrt{N!}}(\hat{a}_{\sigma,p,\ell}^{\dagger})^N\ket{0}\\
    &= \frac{1}{\sqrt{N!}}\left(\int d^2{r}_\bot\, \text{LG}_{p,\ell}(\textbf{r}_\bot;z)\,\hat{a}_{\sigma}^\dagger(\textbf{r}_\bot)\right)^{N}\,\ket{0}.\nonumber
\end{align}

\noindent\emph{$N$ photons having different quantum states:}
\begin{align}\label{eq:NdP}
    \prod^N_{i=1} \ket{1_{\sigma_i,p_i,\ell_i}}:=& \prod^N_{i=1} \hat{a}_{\sigma_i,p_i,\ell_i}^{\dagger}\ket{0}\cr
    =&\int d^2{r_1}_\bot\cdots\int d^2{r_N}_\bot \cr
    &\times\text{LG}_{p_1,\ell_1}({\textbf{r}_1}_\bot;z)\cdots\text{LG}_{p_N,\ell_N}({\textbf{r}_N}_\bot;z)\cr
&\times\left(\hat{a}_{\sigma_1}^\dagger({\textbf{r}_1}_\bot)\cdots\hat{a}_{\sigma_N}^\dagger({\textbf{r}_N}_\bot)\right)\ket{0}.
\end{align}
The difference between these states can be stressed considering the equal-time first-order correlation function, $g^{(1)}(\mathbf{r}_{\bot}):=\bra{\psi}\hat{a}^\dagger(\mathbf{r}_{\bot})\hat{a}(\mathbf{r}_{\bot})\ket{\psi}$, that models intensity measurements. In the case of $N$ photon in the same spatial mode, this yields $g^{(1)}(\mathbf{r}_{\bot})=\abs{\text{LG}_{p,\ell}(\mathbf{r}_{\bot})}^2$, while in the case $\ket{\psi}= \prod^N_{i=1} \ket{1_{\sigma_i,p_i,\ell_i}}$ a straightforward calculation shows that $g^{(1)}(\mathbf{r}_{\bot})=\sum_i\abs{\text{LG}_{p_i,\ell_i}(\mathbf{r}_{\bot})}^2$. Thus, in the first case, an intensity measurement will display the typical shape of an LG mode, while in the second case, one can expect a more `blurred' image arising from the incoherent sum of many different modes.

So far, we have discussed the proper quantum description of multi-photons that can possess various distinct quantum states, such as differing spatial modes, polarisation, or frequency. These photons can possess identical or different quantum states, describing different experiments. Now, we shift our focus to describe the action of optical devices and detectors.

\begin{figure}[!htpb]
    \centering
    \includegraphics[width=1\columnwidth]{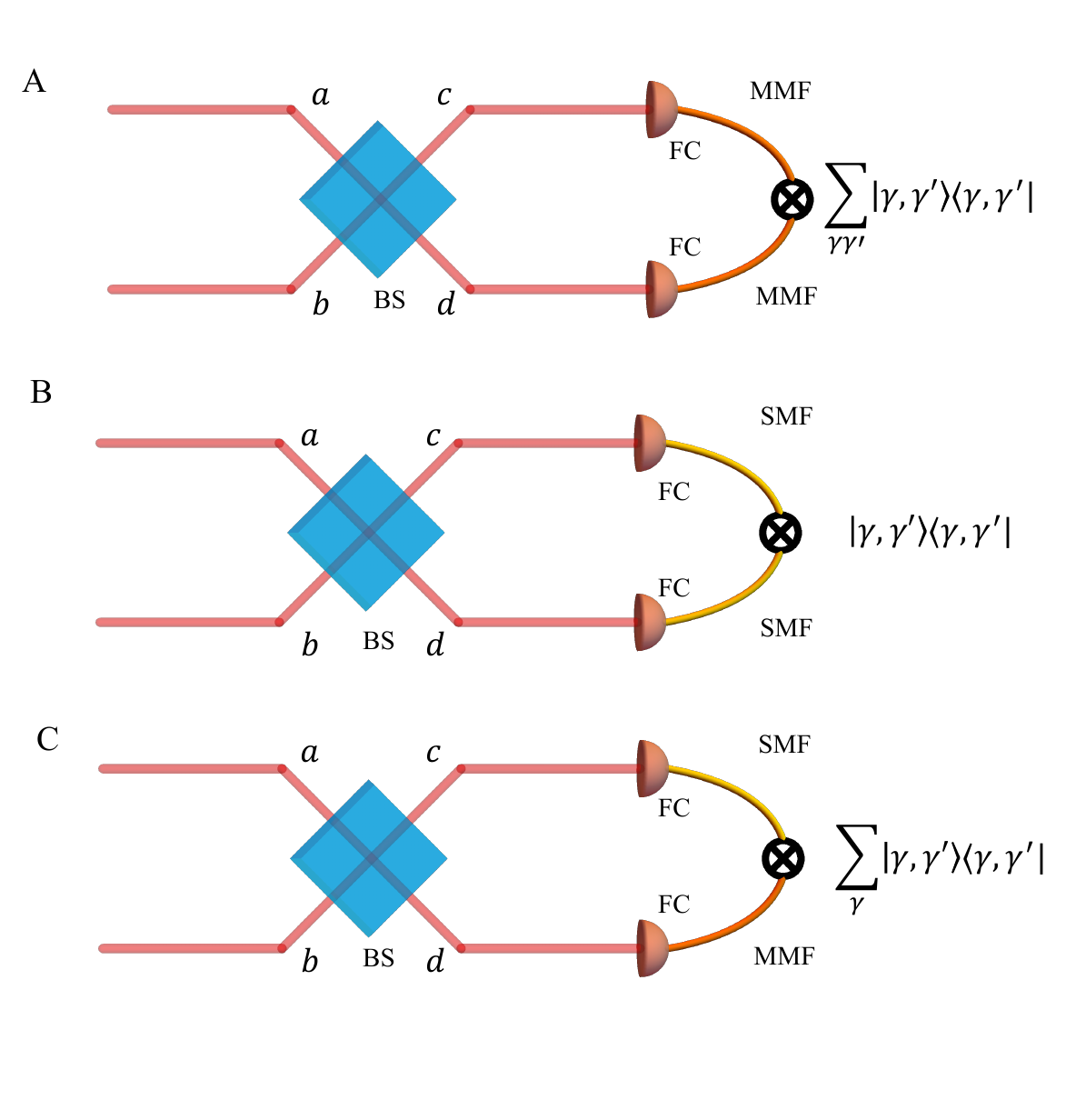}
    \caption{Examples of HOM experimental configurations. Photon pairs enter ports $a$ and $b$ of a beamsplitter, BS, and coincidences counts between the two output paths $c$ and $d$ are collected, typically after coupling to fibres (FC). Each detector can be either a single mode, represented by a single mode fibre (SMF), or a bucket/multimode detector, represented by a multimode fibre (MMF).}
    \label{fig:hom}
\end{figure}

A lossless linear device, e.g. a phase shifter or a beamsplitter, or a waveplate, can be described as a linear transformation mapping the creation operators in a specific mode into a superposition of creation operators in different modes \cite{kok2007linear}:
\begin{eqnarray}
    \hat{a}^\dagger_{\alpha}\rightarrow \sum_\beta u_{\alpha,\beta}\hat{a}^\dagger_{\beta},
\end{eqnarray}
where the coefficients $u_{\alpha,\beta}$ are the entries of a $M\times M$ unitary matrix, with $M$ being the number of modes involved. The unitarity requirement comes from the assumption of the absence of losses, equivalent to the conservation of the number of photons. The indices $\alpha, \,\beta$ can refer to different sets of mode labels.
The simplest example is the phase shifter that simply multiplies the input mode by a phase factor $\hat{a}^{\dagger}_{\alpha}\rightarrow e^{i\theta}\hat{a}^{\dagger}_{\alpha}$. This operation can describe the propagation through a thin slab of material or in free space. However, for long enough propagation ranges, one must take into account that the phase $\theta$ can depend on the spatial mode (so it is at least in part conditioned by the index $\alpha$).
A second example of mode transformation is the beamsplitter, the essential tool for HOM interference. The beamsplitter (BS) is a slab of material tilted with respect to the incident photons at a non-zero angle that can partially reflect and partially transmit light. For a consistent quantum mechanical description of the BS operation, one should consider photons incident on either face of the material. Under the condition that the incidence angles are fixed (with a small uncertainty allowed by the paraxial approximation), the BS acts as a two-input, two-output device. It is also convenient to define with $(\hat{a}^{\dagger}, \hat{b}^{\dagger})$ the vector of input creation operators of photons entering through port $(a,b)$ (see Figure~\ref{fig:hom}) and with $(\hat{c}^{\dagger}, \hat{d}^{\dagger})$ the vector of input creation operators of photons exiting through port $(c,d)$ --as for now we do not specify additional mode indices in the creation operators. Conservation of photon number implies that the BS is described by a transformation of the form~\cite{prasad1987quantum,ou1987relation,fearn1987quantum}:
\begin{eqnarray}
\left(
\begin{array}{c}
  \hat{a}^{\dagger}  \\
  \hat{b}^{\dagger}
\end{array}
\right)
=
\left(
\begin{array}{cc}
  r^*  &  t  \\
  -t^*  &  r 
\end{array}
\right) \left(
\begin{array}{c}
  \hat{c}^{\dagger}  \\
  \hat{d}^{\dagger}
\end{array}
\right),
\end{eqnarray}
with $\abs{r}^2+\abs{t}^2=1$. The factors $r$ and $t$ are associated with the reflectivity and the transmissivity of the device (this can be indeed confirmed by calculating the action of the BS on an ideal laser beam, modelled as a coherent state \cite{ou1987relation}). For incident fields described by paraxial spatial modes, the coefficients $r$ and $t$ are approximately independent of the incident mode except for the OAM modes, since the index $\ell$ changes sign under reflection. Thus, for LG modes
\begin{eqnarray}
\left(
\begin{array}{c}
  \hat{a}^{\dagger}_{\ell,p}  \\
  \hat{b}^{\dagger}_{-\ell,p}
\end{array}
\right)
=
\left(
\begin{array}{cc}
  r^*  &  t  \\
  -t^*  &  r 
\end{array}
\right) \left(
\begin{array}{c}
  \hat{c}^{\dagger}_{-\ell,p}  \\
  \hat{d}^{\dagger}_{\ell,p}
\end{array}
\right).
\end{eqnarray}
The BS is thus the prototypical example of a $2\times2$ unitary operation, i.e. a single qubit gate. Here, we conveniently choose the subscripts $\pm\ell$ in such a way that the output $c$ is associated with $-\ell$ and $d$ with $+\ell$.\newline

Generic quantum processing protocols can be implemented by combining beamsplitters, phase shifters and measuring devices \cite{kok2007linear}. The latter completes the toolkit for HOM interference. As we shall see, the outcome of HOM experiments is strongly affected by the choice of detectors.
The action of a detector determines how quantum states are measured and influences the results of quantum experiments \cite{mandel1995optical}. A bucket detector is a simple example, and in fact, it is insensitive to the polarization state, frequency, or spatial mode of the incoming photon, meaning that it does not discriminate between these degrees of freedom. The action of a single photon bucket detector effectively projects onto a mixed state, represented as:
\begin{align}\label{eq:bucketD}
    \widehat{\Pi}&=\ket{1}\!\bra{1}\cr
    &=\sum_{\sigma}\int d^2{r}_\bot \,\hat{a}_\sigma^\dagger(\textbf{r}_\bot)\ket{0}\!\bra{0}\hat{a}_{\sigma}(\textbf{r}_\bot).
\end{align}
When performing a measurement on a quantum system described by the density matrix $\hat{\rho}_{\ket{\psi}}$, and detecting a single photon without regard to its specific quantum state, the probability of detecting the photon using a bucket detector is given by the following,
\begin{align}\label{eq:bucketDQS}
    \text{Probability}=\text{Trace}\left(\widehat{\Pi}\,\hat{\rho}_{\ket{\psi}}\right),
\end{align}
where $\text{Trace}(.)$ is the trace of the operator.

If the detector is sensitive to a specific spatial mode, e.g. $\text{LG}_{q,m}$, and is capable of detecting only one single photon, then the action of the detector is given by
\begin{eqnarray}\label{eq:LGDetector}
    &{}&\widehat{\Pi}_{q,m}=\ket{1}_{q,m}\!\!\bra{1}\\
    &=&\int d^2{r}_\bot d^2{r}_\bot'\,\text{LG}(\textbf{r}_\bot)\hat{a}^\dagger(\textbf{r}_\bot)\ket{0}\!\bra{0}\hat{a}(\textbf{r}'_\bot) \text{LG}^*(\textbf{r}_\bot').\nonumber
\end{eqnarray}
In this case, the probability of detecting a photon from a physical system of $\hat{\rho}_{\ket{\psi}}$ using a detector that is capable of performing a projective measurement of a desired spatial mode of $\text{LG}_{q,m}$ is given by the following,
\begin{eqnarray}\label{eq:LGDQS}
\text{Probability}=\text{Trace}\left(\widehat{\Pi}_{q,m}\,\widehat{\rho}_{\ket{\psi}}\right).
\end{eqnarray}
With this in mind, we are ready to move on to the description of the HOM effect after different detection conditions.

\section{HOM visibility under different detection configurations.}
After introducing the formalism describing the transformation of electromagnetic field operators by macroscopic lossless media and the effect of projective measurements, the stage is ready to investigate the Hong-Ou-Mandel interference between polarization and frequency-degenerate photons. In this tutorial, we want to stress the effect of the spatial modes involved in the input fields and in the detection stage. The treatment of HOM in the frequency degree of freedom can be found in several references, including Ref.~\cite{ou2007multi}.

\noindent\textit{Input state:} Several choices of input states are possible depending on the experimental scenario. Here, we focus on the simplest case of two photons prepared in a product state where each photon enters the beamsplitter from a different port. We later generalize the results to input correlated photon pairs. Let us assume that the photon entering in port $a$ is in the spatial mode $\psi_1(\mathbf{r}_{\bot})$ and the photon entering port $b$ is in the spatial mode $\psi_2(\mathbf{r}_{\bot})$. This is equivalent to the statement that the input state is of the form $\ket{\psi_0}=\ket{\psi_1}_a\otimes\ket{\psi_2}_b$. The explicit expression of $\ket{\psi_{1,2}}$ depends on the choice of mode decomposition. Let us denote with $\hat{a}^{\dagger}_{\alpha}$ a creation operator for photons entering the port $a$ in mode $\alpha$ (to be specified), and with $\hat{b}^{\dagger}_{\beta}$ a creation operator for photons entering the port $b$ in mode $\beta$. If $\alpha$ and $\beta$ are associated with the transverse position then $\ket{\psi_{1}}=\int dr_{\bot}^2\psi_{1}(\mathbf{r}_{\bot})\hat{a}^{\dagger}_{\mathbf{r}_{\bot}}\ket{0}$ and $\ket{\psi_{2}}=\int dr_{\bot}^2\psi_{2}(\mathbf{r}_{\bot})\hat{b}^{\dagger}_{\mathbf{r}_{\bot}}\ket{0}$. Alternatively, $\alpha$ and $\beta$ can be sets of discrete indices associated with orthonormal sets of spatial modes (e.g. $\ell$, and $p$). For instance, a decomposition in LG modes yields $\ket{\psi_{1}}=\sum_{\ell,p}(\ell,p|\psi_1)\hat{a}^{\dagger}_{\ell,p}\ket{0}$ and $\ket{\psi_{2}}=\sum_{\ell,p}(\ell,p|\psi_2)\hat{b}^{\dagger}_{\ell,p}\ket{0}$. Here we used the notation $(\ell,p|\psi_{1,2}):=\int d^2r_{\bot}\text{LG}^*_{\ell,p}(\mathbf{r}_{\bot})\psi_{1,2}(\mathbf{r}_{\bot})$ to denote the hermitian product in the space of complex functions with integrable square. In the following calculation, we will not specify whether we choose one or the other mode decomposition:
\begin{equation}   \ket{\psi_0}=\ket{\psi_1}_a\otimes\ket{\psi_2}_b=\sum_{\alpha,\beta}\psi_{1,\alpha}\psi_{2,\beta}\hat{a}^{\dagger}_{\alpha}\hat{b}^{\dagger}_{\beta}\ket{0},
\end{equation}
where the summation symbol has to be intended as an integral if the indices $\alpha$ and $\beta$ are continuous. \newline

\noindent\textit{Action of the beamsplitter:} We consider for simplicity a 50:50 beamsplitter (BS) acting on the input modes as follows:
\begin{align}
\hat{a}^{\dagger}_{\alpha}&\rightarrow(\hat{c}^{\dagger}_{\alpha}+\hat{d}^{\dagger}_{\alpha})/\sqrt{2}\cr
\hat{b}^{\dagger}_{\beta}&\rightarrow(\hat{c}^{\dagger}_{\beta}-\hat{d}^{\dagger}_{\beta})/\sqrt{2},
\end{align}
where $c$ and $d$ refer to the output ports of the beamsplitter. Thus, the output state is
\begin{align}
\ket{\psi_{out}}=\sum_{\alpha,\beta}\psi_{1,\alpha}\psi_{2,\beta}\frac{(\hat{c}^{\dagger}_{\alpha}+\hat{d}^{\dagger}_{\alpha})(\hat{c}^{\dagger}_{\beta}-\hat{d}^{\dagger}_{\beta})}{2}\ket{0}.
\label{eq:outstate}
\end{align}\newline

\noindent\textit{Measurement:} Typical HOM experiments consist of measuring the coincidence counts at the output ports $c$ and $d$ of the BS. This can be formalized as the outcome of a simultaneous projective measurement $\hat{\Pi}_{cd}=\hat{\Pi}_c\otimes\hat{\Pi}_d$. Without losing generality, we can model the projection operators in the form 
\begin{equation}
\hat{\Pi}_{c}=\sum_{\gamma}p_{\gamma}\hat{c}^{\dagger}_{\gamma}\ket{0}\bra{0}\hat{c}_{\gamma},
\end{equation}
where, from $\hat{\Pi}^2=\hat{\Pi}$ (the defining property of projection operators), the real coefficients $p_{\gamma}$ must obey $p_\gamma=0,1$. These different choices allow us to distinguish between bucket, few-mode, and single-mode detectors. The projector operator for coincidence measurements is, thus,
\begin{equation}
\hat{\Pi}_{cd}=\sum_{\gamma,\gamma'}p^{(c)}_{\gamma}p^{(d)}_{\gamma'}\hat{c}^{\dagger}_{\gamma}\hat{d}^{\dagger}_{\gamma'}\ket{0}\bra{0}\hat{c}_{\gamma}\hat{d}_{\gamma'},
\end{equation}
whence the measurement output will be 
\begin{align}
    R_{c,d}&:=\text{Tr}[\hat{\rho}_{out}\hat{\Pi}_{cd}]=\text{Tr}[\ket{\psi_{out}}\bra{\psi_{out}}\hat{\Pi}_{cd}]\cr
&=\sum_{\gamma,\gamma'}p^{(c)}_{\gamma}p^{(d)}_{\gamma'}\bra{\psi_{out}}\hat{c}^{\dagger}_{\gamma}\hat{d}^{\dagger}_{\gamma'}\ket{0}\bra{0}\hat{c}_{\gamma}\hat{d}_{\gamma'}\ket{\psi_{out}}\cr
    &=\sum_{\gamma,\gamma'}p^{(c)}_{\gamma}p^{(d)}_{\gamma'}\abs{\bra{0}\hat{c}_{\gamma}\hat{d}_{\gamma'}\ket{\psi_{out}}}^2.
\end{align}
We now focus on the term $\bra{0}\hat{c}_{\gamma}\hat{d}_{\gamma'}\ket{\psi_{out}}$. It is evident that, from Eq.~\eqref{eq:outstate}, only terms with a photon in mode $d$ and a photon in mode $c$ can contribute to $R_{c,d}$. The non-zero contribution is
\begin{align}
\bra{0}\hat{c}_{\gamma}\hat{d}_{\gamma'}\ket{\psi_{out}}=&\frac{1}{2}\sum_{\alpha,\beta}\psi_{1,\alpha}\psi_{2,\beta}\bra{0}\hat{c}_{\gamma}\hat{d}_{\gamma'}\hat{c}^{\dagger}_{\beta}\hat{d}^{\dagger}_{\alpha}\cr&-\hat{c}_{\gamma}\hat{d}_{\gamma'}\hat{c}^{\dagger}_{\alpha}\hat{d}^{\dagger}_{\beta}\ket{0}\cr =&\frac{1}{2}\sum_{\alpha,\beta}\psi_{1,\alpha}\psi_{2,\beta}(\delta_{\gamma,\beta}\delta_{\gamma',\alpha}-\delta_{\gamma,\alpha}\delta_{\gamma',\beta})\cr =&
\frac{1}{2}(\psi_{1,\gamma'}\psi_{2,\gamma}-\psi_{1,\gamma}\psi_{2,\gamma'}).
\end{align}
Substituting, we obtain the general result
\begin{equation}
R_{c,d}=\frac{1}{4}\sum_{\gamma,\gamma'}p^{(c)}_{\gamma}p^{(d)}_{\gamma'}\abs{(\psi_{1,\gamma'}\psi_{2,\gamma}-\psi_{1,\gamma}\psi_{2,\gamma'})}^2,
\label{eq:coincidences_general}
\end{equation}
where is useful to recall that $\psi_{i,\gamma}=(\gamma|\psi_i)$, with $i=1,2$. In the following section, we specialize this result into three scenarios that can be often encountered in experiments

\subsection{Example 1: Two bucket detectors}
If both the detectors in arm $c$ and $d$ are wide-area free space sensors or coupled to multimode fibres, i.e. they can collect a large number of modes, then $\hat{\Pi}_{c,d}$ is characterized by $p^{(c,d)}_{\gamma}=1$ for each mode $\gamma$ that can be detected by the sensors. Hence, the coincidence rate is proportional to 
\begin{align}
    R_{c,d}=&\frac{1}{4}\sum_{\gamma,\gamma'}\abs{(\psi_{1,\gamma'}\psi_{2,\gamma}-\psi_{1,\gamma}\psi_{2,\gamma'})}^2\cr=&\frac{1}{2}\biggr(\sum_{\gamma}\abs{\psi_{1,\gamma}}^2\sum_{\gamma'}\abs{\psi_{2,\gamma'}}^2-\cr&\sum_{\gamma}\psi^*_{1,\gamma}\psi_{2,\gamma}\sum_{\gamma'}\psi^*_{2,\gamma'}\psi_{1,\gamma}\biggr).
\end{align}
This expression can be further simplified if all the modes contained in the expansion of $\psi_{1,2}$ are detectable, i.e. $\sum_{\gamma}\abs{\psi_{(1,2),\gamma}}^2=1$. Hence, one can see that the coincidence rate yields a direct measure of the fidelity between the two input modes
\begin{align}
R_{c,d}=&\frac{1}{2}(1-\abs{(\psi_1|\psi_2)}^2).
\end{align}

\subsection{Example 2: Single-mode detectors}
It is also common to use single-mode detectors such as single-mode fibres (SMF), which effectively project on (approximately) Gaussian modes, or detector arrays such as time stamping cameras or EMCCDs \cite{devaux2020imaging,zhang2021high,gao2022high}. These scenarios can be modelled assuming that $p^{(c,d)}_{\gamma}=\delta_{\gamma, \eta_{(c,d)}}$, where, in general, $\eta_{c}\neq\eta_{d}$ (for instance one can consider pixel modes corresponding to different transverse positions or two single-mode fibres coupled via different objective lenses). The coincidence rate is
\begin{align}
R_{c,d}=&\frac{1}{4}\abs{(\psi_{1,\eta_c}\psi_{2,\eta_d}-\psi_{1,\eta_d}\psi_{2,\eta_c})}^2,
\label{eq:coinc_singlemode1}
\end{align}
where $R_{c,d}\rightarrow 0$ for $\eta_{d}\rightarrow\eta_{c}$, that is, one can get maximum HOM visibility if the projection modes are identical and independently of the input modes. This is in contrast with the bucket detection case, where there is an overlap between input modes that determines the HOM visibility. 

Note also that Eq.~\eqref{eq:coinc_singlemode1} carries information between the relative phase structure between $\psi_1$ and $\psi_2$. This has been exploited in holographic experiments \cite{chrapkiewicz2016hologram,thekkadath2023intensity}.

\subsection{Example 3: Single-mode detector and bucket detector}
An interesting hybrid configuration can be achieved when one of the detectors is a single mode and the second detector is a bucket detector. For instance, if the single mode detector is a single photon-resolving camera, this is the typical configuration of ghost imaging experiments; however, in our scenario, we are considering mixing signal and idler in a BS first. The detection is modelled by $p^{(c)}_\gamma=\delta_{\gamma,\eta}$ and $p^{(d)}_\gamma=1,\, \forall \gamma$. From Eq. \eqref{eq:coincidences_general} we obtain
\begin{align}
R_{c,d}=&\frac{1}{4}\sum_{\gamma}\abs{(\psi_{1,\gamma}\psi_{2,\eta}-\psi_{2,\gamma}\psi_{1,\eta})}^2\cr
=&\frac{1}{4}[\abs{\psi_{1,\eta}}^2+\abs{\psi_{2,\eta}}^2-\cr&(\sum_{\gamma}\psi_{1,\gamma}\psi^*_{2,\gamma}\psi_{2\eta}\psi^*_{1,\eta}+c.c.)].
\end{align}\newline
The last expression can be made more explicit by recalling that $\psi_{1,\gamma}=(\gamma|\psi_1)$ and using the completeness relationship $\sum_\gamma |\gamma)(\gamma|=\hat{1}$, we obtain
\begin{align}
R_{c,d}=&\frac{1}{4}\biggr\{\abs{(\eta|\psi_1)}^2+\abs{(\eta|\psi_2)}^2-\cr&\bigr[(\psi_2|\psi_1)(\eta|\psi_2)(\psi_1|\eta)+c.c\bigr]\biggr\}.
\end{align}

\subsection{Generalization to correlated input states}
Hitherto, we have considered the case of input states that can be described as product states of two photons entering either port of the BS \cite{walborn2010spatial}. In many experiments, one encounters a different scenario where the input state is non-separable in the modes of light. This is typically the case when SPDC states generated in free space are directly input to the BS. A more general description of HOM interference is obtained assuming as input state $\ket{\psi_{0}}=\sum_{\alpha,\beta}\psi_{\alpha,\beta}\hat{a}^{\dagger}_{\alpha}\hat{b}^{\dagger}_{\beta}\ket{0}$, where the assumptions are that the two-photon state is pure and one photon is always entering from port $a$ and the other from port $b$. This condition can be realized experimentally by exploiting anti-correlations in momentum --e.g. using a knife edge mirror placed in the far field of the crystal, or in polarization (for Type-II SPDC sources) --using a polarizing beamsplitter-- to deterministically send the idler photon in path $a$ and the signal photon in path $b$. Following the steps for the calculation of HOM interference, one has that the coincidence rate is given by
\begin{align}
    R_{c,d}=\frac{1}{2}\sum_{\gamma,\gamma'}p_\gamma p_\gamma'\abs{(\psi_{\gamma,\gamma'}-\psi_{\gamma',\gamma})}^2.
\end{align}
For arbitrary transmissivity (unbalanced BS), one gets:

\begin{align}
    R_{c,d}=\sum_{\gamma,\gamma'}p_\gamma p_\gamma'\abs{(\abs{r}^2\psi_{\gamma,\gamma'}-\abs{t}^2\psi_{\gamma',\gamma})}^2,
\end{align}
which highlights how imperfect visibility can also be due to differences in the reflectivity and transmissivity of the BS.

\section{Explicit formulae for Laguerre-Gauss modes basis}
\begin{figure}
    \centering
\includegraphics[width=1\columnwidth]{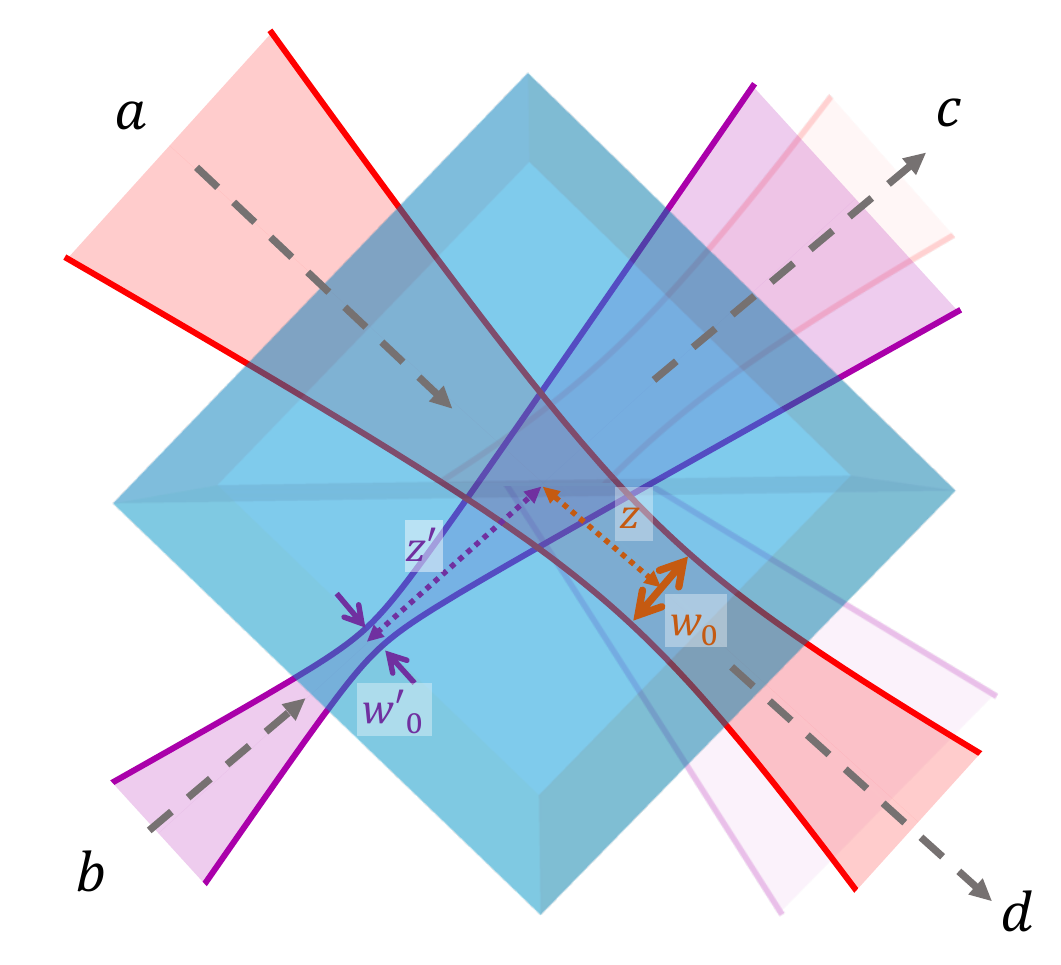}
    \caption{Illustration of HOM interference geometry for input photons having different mode parameters.}
    \label{fig:hommodes}
\end{figure}
From the previous sections, it is evident how the coincidence rate in HOM experiments requires the evaluation of inner products between spatial modes $(\psi_1|\psi_2)$. The completeness and orthogonality of the  Laguerre-Gauss modes allows to always expand $\psi_{1,2}$ as an infinite series of LG functions. The inner product will thus become a superposition of inner products between different LG modes. Two approaches can be used:\newline

\noindent\textit{(1)} $\psi_1$ and $\psi_2$ are decomposed in the same LG basis (same choice of waist $w_0$ and propagation distance $z$). In this case, if $\psi_{1,2}=\sum_{\ell,p}A^{(1,2)}_{\ell,p}\text{LG}_{\ell,p}$,  $\psi_1(\mathbf{r}_{\bot}) = \text{LG}_{l,p}(\mathbf{r}_{\bot}; z)$, then 
\begin{align}(\psi_1|\psi_2)&=\sum_{\ell,\ell',p,p'}A^{*(1)}_{\ell,p}A^{(2)}_{\ell',p'}(\ell,p|\ell',p')\cr&=\sum_{\ell,p}A^{*(1)}_{\ell,p}A^{(2)}_{\ell,p},
\label{eq:LGprod_dec}
\end{align} 
where we used $(\ell,p|\ell',p')=\delta_{\ell,\ell'}\delta_{p,p'}$.\newline

\noindent\textit{(2)} The modes $\psi_{1,2}$ may be more conveniently expanded in LG functions with different $w_0$ and $z$ for modes 1 and 2. Indeed, the number of relevant coefficients $A_{\ell,p}$ can be reduced significantly by looking for an optimal decomposition waist (a nice example for Hypergeometric-Gaussian modes is given in Ref.~\cite{vallone2017role}). This situation can be of interest when the modes incident on the beamsplitter have different radii and wavefront curvature, as illustrated in Fig. \ref{fig:hommodes}. In these scenarios, the product $(\ell,p|\ell',p')$ in Eq.~\eqref{eq:LGprod_dec}
should be replaced by the complex-valued integral
\begin{align}
    \mathcal{I}_{\ell,\ell',p,p'}:= \int^{2\pi}_{0} d \phi \int^{\infty}_{0} \rho d\rho  \text{LG}^{*}_{\ell,p}(\mu\rho, \phi; z) \text{LG}_{\ell',p'}(\mu'\rho, \phi; z'),
\end{align}
where 
\begin{align}
    \mu &= \frac{2}{w(z)^{2}}.
\end{align}
We hereafter introduce $w'(z')$ and $R'(z')$ to indicate, respectively, the beam waist and the radius of curvature of the beam at propagation distance $z'$ and having waist $w^{'}_o$ and, consequently, Rayleigh range $z'_o$. Thus, $\mu'$ is a coefficient that depends on $w'(z')$. In experiments, if the distances and the characteristics of each optical element before the BS are known, the values of $z,\, z',\, w_o$ and $w'_o$ can be calculated from the ABCD matrix of the setup (see, e.g. Refs.  \cite{siegman1986lasers,yariv2007photonics,saleh2019fundamentals}).
From Eq. \eqref{eq:LG}, the explicit formula of the LG modes can be applied directly to the integral. The azimuthal part yields a $\delta_{l,l'}2\pi$ factor. We, therefore, evaluate the case where $l=l'$.  By performing the substitution $x = \rho^2$, the integral can be put into the following form 
\begin{align}
    \mathcal{I}_{\ell,\ell,p,p'} &= A \int_{0}^{\infty} dx x^{|\ell|} e^{-\sigma x} L^{|\ell|}_{p}(\mu x) L^{|\ell|}_{p'}(\mu' x),
\end{align}
where 
\begin{eqnarray}
A =& 2^{|\ell|-1} \sqrt{\frac{2\,p!}{\pi(p+|\ell|)!}} \sqrt{\frac{2\,p'!}{\pi(p'+|\ell|)!}} \left(\frac{1}{w(z)w'(z')}\right)e^{i\Delta\psi_G}
\cr
\sigma=&\frac{1}{w(z)^{2}} + \frac{1}{w'(z')^{2}}+ i \frac{k}{2} \left(-\frac{1}{R(z)}  + \frac{1}{R'(z')}\right), 
\end{eqnarray}
with $\Delta\psi_G:=(2p' + |\ell'| + 1)\arctan(\frac{z'}{z'_{o}})-(2p + |\ell| + 1)\arctan(\frac{z}{z_{o}})$ is the difference between the accumulated geometric phases.
 
The integral has an analytical solution expressed in terms of the Hypergeometric Function $_2F_1$ \cite{gradshteyn2014table}
\begin{align}
    (\ell,p|\ell',p') = A \frac{\Gamma(p+p'+ |\ell| + 1)}{p!p'!} \frac{(\sigma-\mu')^{p'}(\sigma - \mu)^{p}}{\sigma^{p+p'+|\ell|+1}} \nonumber \\ \times _2F_1\left(-p,-p';-p-p'-|\ell|; \frac{\sigma(\sigma - \mu' - \mu)}{(\sigma-\mu')(\sigma-\mu)}\right) \label{eq:LG_overlap_analytical} 
\end{align}

\begin{figure}[!htbp]
    \centering
    \includegraphics[width=\columnwidth]{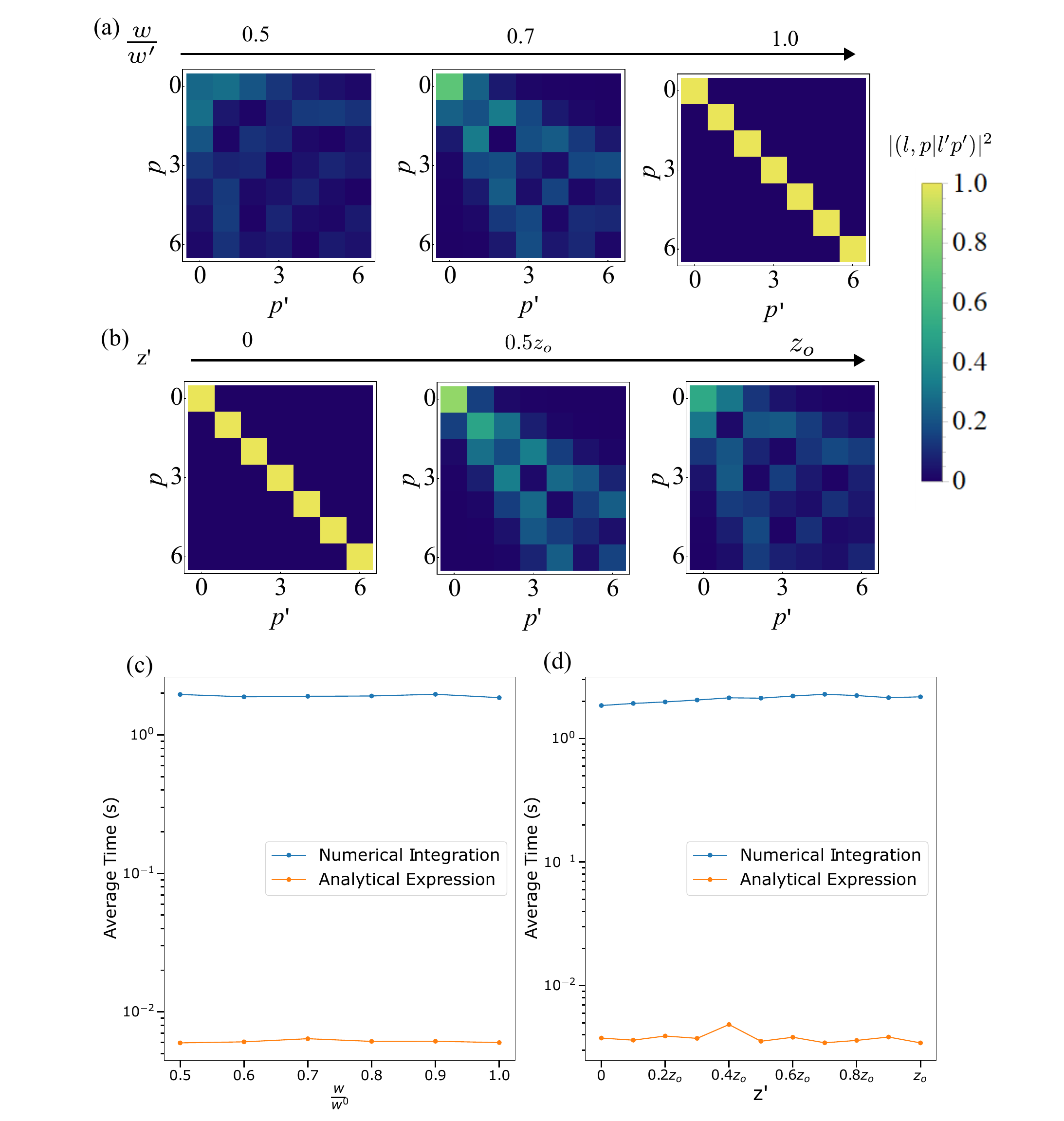}
    \caption{Evaluation of $|(l,p|l'p')|^2$ with $l=l'=2$ and for different combinations of $p,p' \in [0,6]$. We demonstrate how the overlap between two beams changes (a) when the ratio of waists and (b) when the relative propagation distance between both beams changes. In both cases, the overlap between identical LG modes falls significantly, and we may observe ``crosstalk" with other LG modes. We also report in (c) and (d) the average computational time of 100 evaluations using numerical integration and the analytical expression of the overlap given by Eq.~\eqref{eq:LG_overlap_analytical}. We repeat the evaluations in (a) and (b), respectively, evaluating in (c) the relative change in beam waist and in (d) the relative change in propagation distance. Both computations were performed in Python on an AMD Ryzen 4500U @ 2.38 GHz CPU, and we use \texttt{SciPy}'s quadrature method to perform the numerical integration~\cite{virtanen2020scipy}.}
    \label{fig:crosstalks}
\end{figure}

We plot the value of $|(l,p|l'p')|^2$ between different combinations of LG modes in Figs.~\ref{fig:crosstalks} (a) and (b) (recall that this is associated with the HOM visibility when two bucket detectors are used). The usual orthogonality of modes falls off as we vary the beam waist and relative propagation distance between the beams, and ``crosstalk" with other, non-orthogonal LG modes can be observed. We note that experimentally, situations can be achieved where $z,z'\ll z_0$ with careful alignment. Slight mismatches due to experimental imperfection in the waist parameters can often be met and severely affect the HOM visibility. The analytical formula of Eq.~\eqref{eq:LG_overlap_analytical} can further be used to explore different, more unconventional experimental scenarios.

\section{Applications} 

\subsection{Complex quantum networks with spatial modes}
\begin{figure}[!htbp]
    \centering
    \includegraphics[width=\columnwidth]{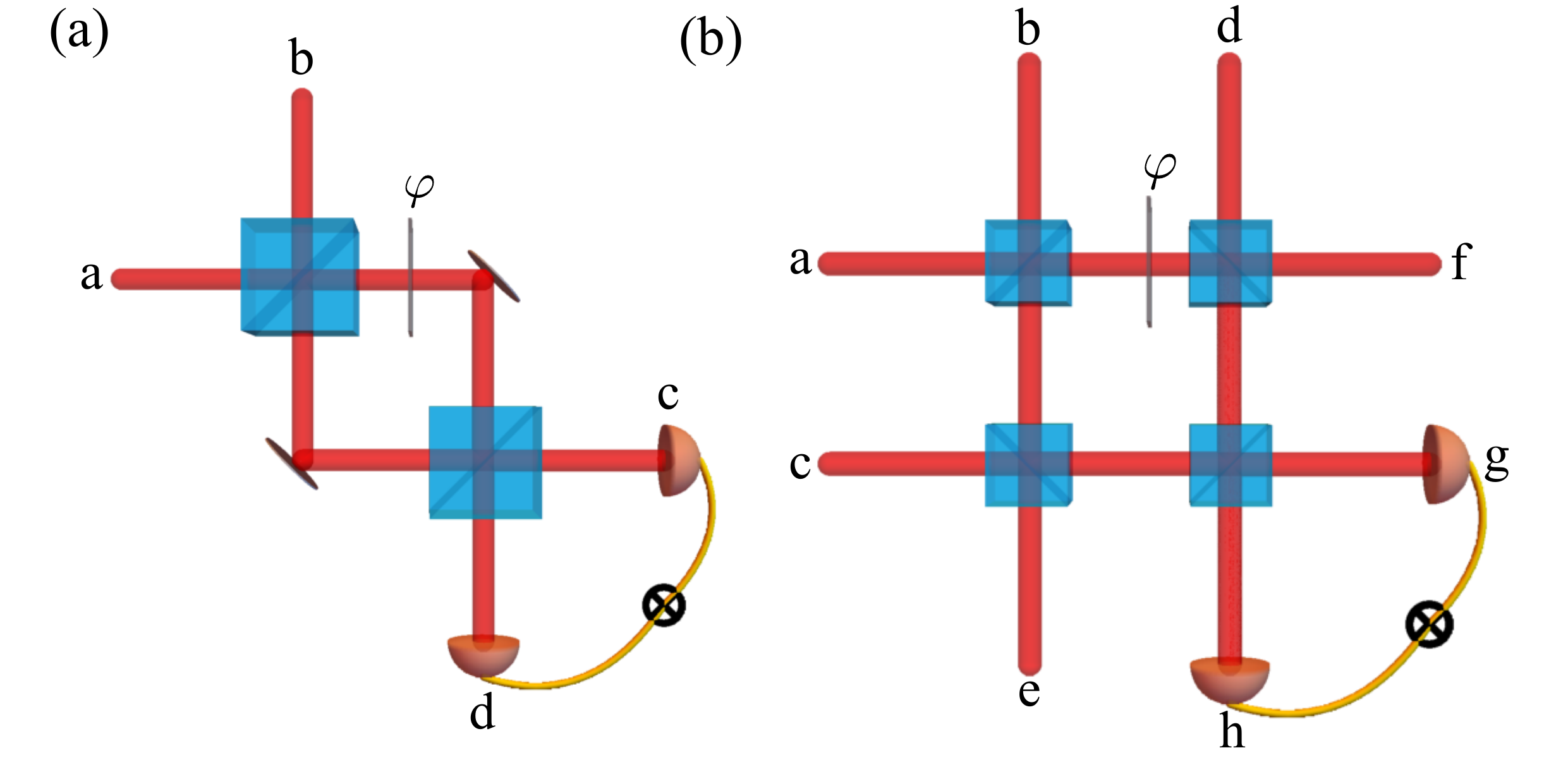}
    \caption{Example experimental scenarios involving (a) two-photon interference with an incoming state $\ket{\psi_{in}}=\ket{1_a}\ket{1_b}$ and (b) multi-photon interference with $\ket{\psi_{in}}=\ket{1_a}\ket{1_b}\ket{1_c}\ket{1_d}$. The indices refer to different input ports to the BSs. The accumulated Gouy phase difference between possible paths can be modelled as a phase-shifting element that encodes a phase $\varphi$. In both cases, the outgoing state and coincidence rate depend on $\varphi$, implicating single-photon interference effects in the outcome of both experiments. }
    \label{fig:photonInterference}
\end{figure}

Computing the results of a linear network requires the precise evaluation of single- and multi-photon interference effects, predominantly the HOM interference as described here. Conventionally, single-photon interference can also be described in a simple way by taking the local phases of photonic modes into account. This scenario becomes more complex, as spatial modes acquire a propagation-dependent phase that depends on the order of the spatial mode -- namely the accumulated Gouy phase \cite{erden1997accumulated, arai2013accumulated}. The difference between the accumulated Gouy phase in different paths of the optical network leads to a single-photon interference effect, which has been shown to be useful for sorting spatial modes \cite{zhou2017sorting, gu2018gouy}. They also impact the outcome of two-photon experiments, even in rather simple experimental scenarios, as we show in Fig.~\ref{fig:photonInterference}.

\subsection{Automated Search for Quantum Optical Experiments}
The task of designing new quantum optical experiments is exacerbated by a gigantic search space for possible configurations of optical elements. Recently, scientists have turned to artificial intelligence to navigate this vast search space~\cite{krenn2020computer, krenn2023artificial}, yielding many novel quantum experimental designs~\cite{krenn2016automated,knott2016search, ruiz2023digital,krenn2023digital}. Here, an algorithm chooses an experiment -- a point in the search space -- and computes its properties with a physical simulator. The results from the simulator are used to compute an objective function (loss function), which acts as a quality measure to help the algorithm to choose a different, hopefully better experiment. The crucial aspect is the physical simulator, which needs to be reliable and fast.

Developing analytical equations for experimental systems, such the formula in Eq.\eqref{eq:LG_overlap_analytical}, will allow us to expand these computational exploration strategies to the domain of quantum states with complex spatial modes.

As a demonstration of the advantage, we can directly compare the numerical approach with the evaluation of the explicit formula for the scalar product $(\ell,p|\ell',p')$. In Figs.~\ref{fig:crosstalks} (c) and (d), we highlight the compute time of the analytical expression over numerically evaluating the overlap. This speed-up will advance the search for quantum optical setups exploiting structured light physics for tasks in precise timing measurements~\cite{lyons2018fast}, and for the generation of high-dimensional N00N states \cite{dowling2008quantum}. N00N states are N-dimensional, entangled states wherein N photons are bunched into multiple modes of a DOF, such as path~\cite{mitchell2004super, walther2004broglie, nagata2007beating, afek2010high}, polarization~\cite{slussarenko2017unconditional, kuzmich1998sub}, and OAM ~\cite{hiekkamaki2022observation, hiekkamaki2021high, d2013photonic}; this corresponds to maximal HOM visibility. They exhibit N-fold sensitivity to phase- and rotational- displacements, depending on the DOF being coupled, making them particularly attractive for quantum metrology tasks.

\begin{funding}
T.J., A.D. and E.K. acknowledge the support of the Canada Research Chairs (CRC), the Max Planck-University of Ottawa Center for Extreme and Quantum Photonics, the Qeyssat User INvestigation Team (QUINT) and the Alliance for Research and Applications of Quantum Network Entanglement (ARAQNE) Alliance Consortia Quantum grants. X. G. acknowledges support from the Alexander von Humboldt Foundation. M.K. acknowledges support by the European Research Council (ERC) under the European Union’s Horizon Europe research and innovation programme (ERC-2024-STG, 101165179, ArtDisQ).
\end{funding}

\begin{authorcontributions}
\emph{All authors contributed to initiating the work and drafting the manuscript.}

\end{authorcontributions}

\begin{conflictofinterest}
\emph{Authors state no conflict of interest.}
\end{conflictofinterest}

\begin{dataavailabilitystatement}
\emph{Upon a reasonable request, the code that compares the evaluation of the Laguerre-Gauss overlap can be shared.}

\end{dataavailabilitystatement}

\bibliographystyle{ieeetr}
\bibliography{bibliography}

\end{document}